\documentclass[aps,prl,twocolumn,superscriptaddress]{revtex4-1}

\usepackage[pdftex]{graphicx}
\usepackage[pdftex,bookmarks=true,bookmarksopen,bookmarksnumbered,
                colorlinks,
                linkcolor=blue,
                citecolor=blue]{hyperref}
\usepackage{dcolumn}
\usepackage{bm}
\usepackage{braket,amsmath}

\begin{document}


\title{Coherent Control of a Single Silicon-29 Nuclear Spin Qubit}

\author{Jarryd J. Pla}
\altaffiliation[Present address: ]{London Centre for Nanotechnology, University College London, 17-19 Gordon Street, London, WC1H 0AH, United Kingdom}
\email{j.pla@ucl.ac.uk}
\author{Fahd A. Mohiyaddin}
\author{Kuan Y. Tan}
\author{Juan P Dehollain}

\affiliation{Centre Excellence for Quantum Computation \& Communication Technology}
\affiliation{School of Electrical Engineering \& Telecommunications, UNSW Australia, Sydney NSW 2052 Australia}
\author{Rajib Rahman}
\author{Gerhard Klimeck}
\affiliation{Network for Computational Nanotechnology, Purdue University, West Lafayette, IN 47907, U.S.A.}
\author{David N. Jamieson}
\affiliation{Centre Excellence for Quantum Computation \& Communication Technology}
\affiliation{School of Physics, University of Melbourne, Melbourne VIC 3010 Australia}
\author{Andrew S. Dzurak}
\author{Andrea Morello}
\email{a.morello@unsw.edu.au}
\affiliation{Centre Excellence for Quantum Computation \& Communication Technology}
\affiliation{School of Electrical Engineering \& Telecommunications, UNSW Australia, Sydney NSW 2052 Australia}

\date{\today}

\begin{abstract}
Magnetic fluctuations caused by the nuclear spins of a host crystal are often the leading source of decoherence for many types of solid-state spin qubit. In group-IV materials, the spin-bearing nuclei are sufficiently rare that it is possible to identify and control individual host nuclear spins. This work presents the first experimental detection and manipulation of a single $^{29}$Si nuclear spin. The quantum non-demolition (QND) single-shot readout of the spin is demonstrated, and a Hahn echo measurement reveals a coherence time of $T_2 = 6.3(7)$~ms -- in excellent agreement with bulk experiments. Atomistic modeling combined with extracted experimental parameters provides possible lattice sites for the $^{29}$Si atom under investigation. These results demonstrate that single $^{29}$Si nuclear spins could serve as a valuable resource in a silicon spin-based quantum computer.

\end{abstract}

\maketitle

The presence of non-zero nuclear spins in a host crystal lattice is known to induce decoherence in a central spin qubit through mechanisms such as spectral diffusion~\cite{Klauder1962_PR}. This ``nuclear bath'' is the primary source of decoherence for $^{31}$P electron and nuclear spin qubits in silicon~\cite{Pla2012_N, Pla2013_N}, nitrogen-vacancy (NV) centers in diamond~\cite{Hanson2008_S}, as well as for GaAs-based quantum dot spin qubits~\cite{Coish2005_PRB,Yao2006_PRB}. However, for semiconductors composed of majority spin-zero isotopes (such as silicon and carbon), the low abundance of spin-carrying nuclei allows to resolve the hyperfine couplings of individual nuclei with a central electronic spin, permitting the detection and manipulation of single nuclear spins. This has led to the demonstration of a quantum register for the spin of a NV center in diamond, where the electronic spin state can be stored in individual nuclei~\cite{Dutt2007_S} and read out in single shot~\cite{Robledo2011_N}. Quantum error correction protocols have been implemented within these nuclear spin registers~\cite{Waldherr2014_N, Taminiau2014_N}, showing their potential to implement surface-code based quantum computing architectures~\cite{Nickerson2013_NC}. Natural silicon contains a 4.7\% abundance of the spin-carrying ($I=1/2$) $^{29}$Si isotope which, in combination with a localized electron spin, could in principle be used as quantum register or ancilla qubit equivalently to $^{13}$C in NV-diamond. In addition, the $^{29}$Si nuclear spin has itself been championed as a quantum bit in an ``all-silicon'' quantum computer~\cite{Ladd2002_PRL, Itoh2005_SSC}.

\begin{figure}[t!]
\includegraphics[width=1\columnwidth]{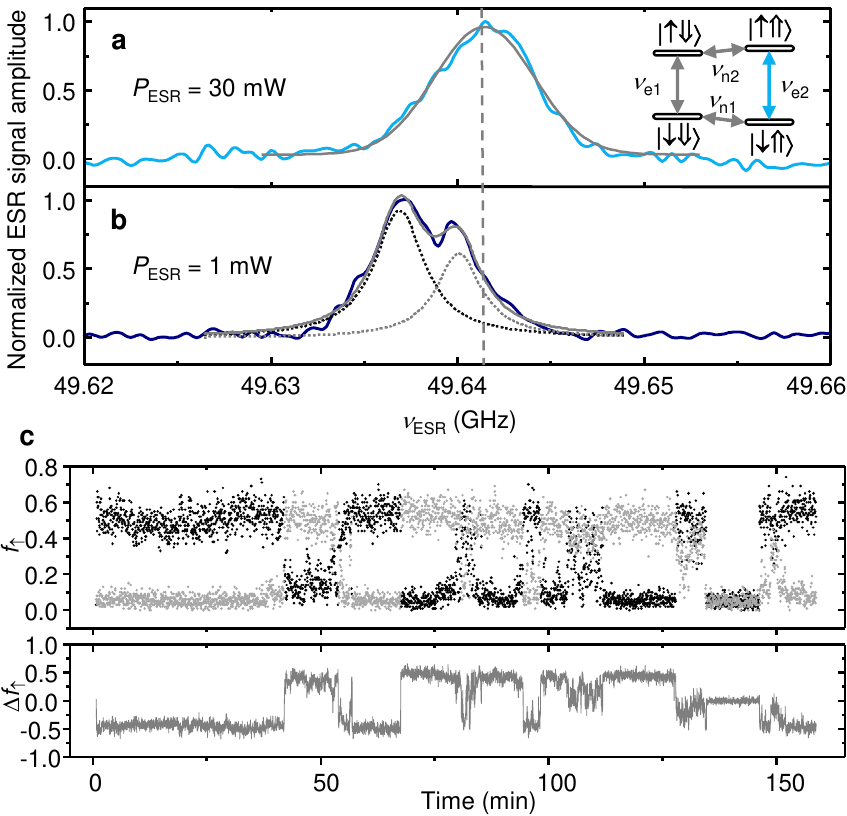}
\caption{\label{fig:figure1} (Color online) ESR scans at the electron spin transition corresponding to the $^{31}$P nuclear $\ket{\Uparrow}$ state ($\nu_{\rm e2}$), performed using microwave powers of (\textbf{a}) $P_{\rm ESR} = 30$~mW and (\textbf{b}) $P_{\rm ESR} = 1$~mW. The data in \textbf{a} is fit with a Gaussian lineshape (gray line). The low-power peak in \textbf{b} displays a splitting of $\sim 2.2$~MHz and is fit with a double-Lorentzian curve (gray line is the sum of the dashed lines). Inset of \textbf{a}: energy level diagram of the $^{31}$P donor system. The $^{29}$Si experiments are performed around the $\nu_{\rm e2}$ resonance. (\textbf{c}) Single-shot readout of a $^{29}$Si nuclear spin. Quantum jumps of the nuclear spin occur on minute-long timescales, with no clear preference for the orientation. Bottom panel: difference in the spin-up fraction $\Delta f_{\uparrow}$ from measurements on the left and right sides of the split $\nu_{\rm e2}$ resonance (shown individually in the top panel).}
\end{figure}

Here we present the first experimental demonstration of single-shot readout, coherent control, and measurement of the coherence properties of an individual $^{29}$Si nuclear spin in natural Si. All measurements were performed with a magnetic field $B_0 = 1.77$~T, in a dilution refrigerator with electron temperature $T_{\rm el} \approx 250$~mK. This work follows from previous experiments where the electron~\cite{Pla2012_N} and nuclear~\cite{Pla2013_N} spins of a single $^{31}$P donor were detected using a compact nano-scale device~\cite{Morello2009_PRB} consisting of ion-implanted phosphorus donors~\cite{Jamieson2005_APL}, tunnel-coupled to a silicon MOS single-electron transistor (SET)~\cite{Angus2007_NL}. Spin control was achieved through microwave and RF excitations generated by an on-chip broadband transmission line~\cite{Dehollain2013_Na}. The $^{31}$P donor in silicon represents a two-qubit system, with an electron spin ($S = 1/2$) bound at cryogenic temperatures to a nuclear spin ($I = 1/2$). The eigenstates of this system are displayed as an inset to Fig.~\ref{fig:figure1}a -- with thin arrows representing the spin state of the electron ($\uparrow,\downarrow$) and thick the nucleus ($\Uparrow,\Downarrow$). There are two electron spin resonance (ESR) frequencies $\nu_{\rm e1,2}$, and two $^{31}$P nuclear magnetic resonance (NMR) frequencies $\nu_{\rm n1,2}$.

The detection of a single $^{29}$Si spin was achieved by first performing an ESR experiment about one of the $^{31}$P hyperfine peaks. We chose the transition corresponding to the $\ket{\Uparrow}$ state, i.e. $\nu_{\rm e2}$, since the nuclear spin is predominantly polarized here as a result of the differing $\ket{\Uparrow}$ and $\ket{\Downarrow}$ nuclear spin relaxation mechanisms~\cite{Pla2013_N}. The ESR experiment involves using the SET to monitor the induced electron spin-up fraction $f_{\uparrow}$ in response to a microwave excitation with varying frequency $\nu_{\rm ESR}$, resulting in the spectrum of Fig.~\ref{fig:figure1}a. The line-shape is well described by a Gaussian with full-width-at-half-maximum (FWHM) $\sim 7$~MHz (or $250~\mu$T) at the largest applied ESR power $P_{\rm ESR} \approx 30$~mW. This figure corresponds to the bulk value for the inhomogeneous broadening caused by the $^{29}$Si nuclear spin bath~\cite{Tyryshkin2003_PRB}. From the measured Rabi frequency at this power~\cite{Pla2012_N} we extract $B_1 \approx 120$~$\mu$T, confirming that power broadening does not occur here. However, by further reducing the excitation power to 1~mW ($B_1 \approx 30$~$\mu$T) the ESR line splits in two, and shifts to lower frequency (Fig.~\ref{fig:figure1}b). A double-Lorentzian fit best captures the shape of the line and yields a FWHM~$\approx 3$~MHz for both peaks, with the center frequency decreasing by 3~MHz with respect to Fig.~\ref{fig:figure1}a. Overall, the observed low-power behavior indicates a polarization and a narrowing of the $^{29}$Si nuclear bath. The behavior is reproducible over several measurements, and does not depend on the direction of the frequency sweep. The microscopic origin of this phenomenon is currently not understood. It is not consistent with the standard Overhauser effect, where excitation of the electron spin to the $\ket{\uparrow}$ state, in combination with a fast electron-nuclear spin-conserving relaxation channel $\ket{\uparrow \Downarrow} \rightarrow \ket{\downarrow \Uparrow}$ results in a predominant $\ket{\Uparrow}$ bath polarization. The line shift to lower frequencies indicates instead a $\ket{\Downarrow}$ polarization, since $^{29}$Si has a negative gyromagnetic ratio $\gamma_{\rm Si} = -8.458$~MHz/T (Ref.~\citenum{Hale1969I_PR}). Several papers have discussed nuclear polarization with anomalous direction, but under conditions that do not apply to our experiment~\cite{Laird2007_PRL,Rudner2007_PRL,Rudner2011_PRB,Urbaszek2013_RMP,Yang2013_PRB}. The line shift and narrowing occurs at low power, when $\gamma_e B_1 \ll \mathrm{FWHM}$, and the resonance is measured through counting single-shot electron spin readout events.  Therefore the experiment effectively constitutes a projective measurement of the nuclear bath state, which can result in a narrowed bath distribution~\cite{Cappellaro2012_PRA}. However, the shift to lower frequencies remains unexplained.

The splitting of the $\nu_{e2}$ line indicates the presence of a single $^{29}$Si nuclear spin, strongly hyperfine-coupled to the donor-bound electron. This allows us to read the $^{29}$Si spin state in the same way as the $^{31}$P spin \cite{Pla2013_N}. Here we apply adiabatic frequency sweeps~\cite{Laucht2014_APL} over the first half of the $\nu_{\rm e2}$ resonance, i.e. from far-detuned to a point mid-way between the two peaks. After each passage we acquire a single-shot measurement of the electron spin to obtain $f_{\uparrow}$. The process is then repeated on the second half of the hyperfine-split $\nu_{\rm e2}$ peak. We observe clear ``quantum jumps'' (Fig.~\ref{fig:figure1}c), providing strong evidence that the splitting does indeed originate from a single spin coupled to the electron. Occasionally, both sides of the split peak produces no resonance, indicating that the $^{31}$P nuclear spin has flipped to $\ket{\Downarrow}$. We therefore periodically measure the state of the donor nuclear spin and initialize it in the $\ket{\Uparrow}$ state if it has flipped~\cite{supp}.

\begin{figure}[t!]
\includegraphics[width=1\columnwidth]{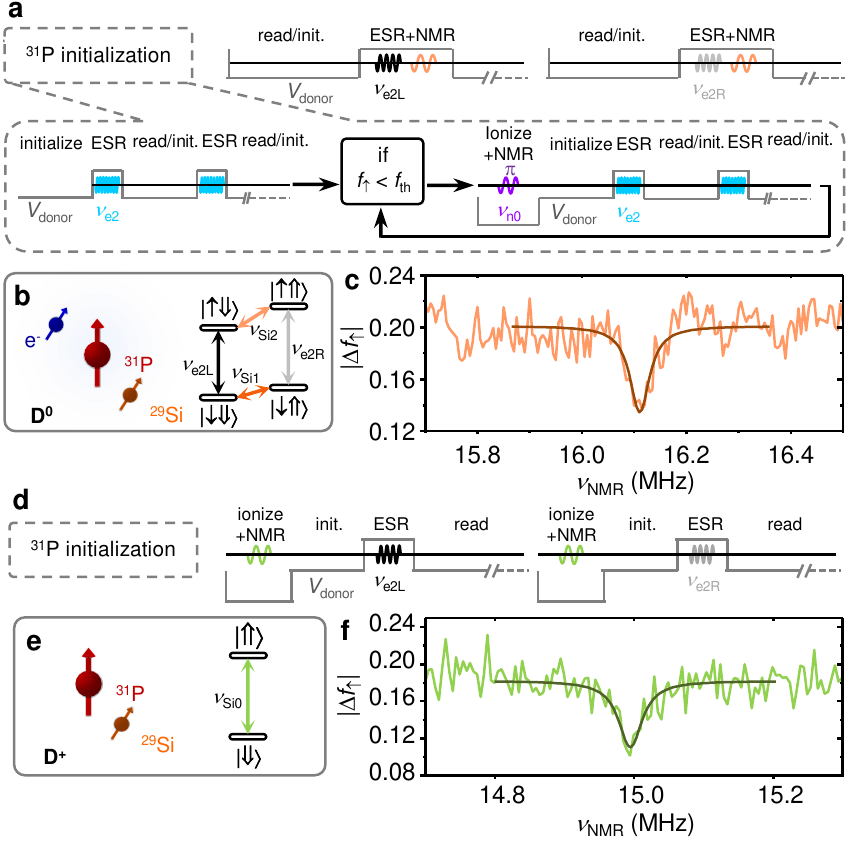}
\caption{\label{fig:figure2} (Color online) (\textbf{a}) Pulse sequence, adapted from Ref.~\citenum{Pla2013_N}, for observing the electron $\ket{\uparrow}$ $^{29}$Si NMR transition $\nu_{\rm Si2}$. Here $V_{\rm donor}$ represents a series of voltage pulses applied to an electrode above the donor to control the electrochemical potential of the bound electron. Preceding the NMR experiment is an initialization of the $^{31}$P nuclear spin~\cite{supp}. (\textbf{b}) Energy level diagram of the neutral (D$^0$) $^{29}$Si:$^{31}$P system, with corresponding ESR and NMR transitions, assuming a fixed $^{31}$P nuclear spin state $\ket{\Uparrow}$ ($m_P = +1/2$). (\textbf{c}) Absolute electron spin-up fraction difference $|\Delta f_{\uparrow}|$ as a function of the NMR frequency $\nu_{\rm NMR}$, for the $^{29}$Si spin with a neutral donor and $m_e = m_P = +1/2$. The resonance is best fit by a Lorentzian function, indicating possible power broadening. (\textbf{d}) Pulse sequence, and (\textbf{e}) energy level diagram for the ionized donor (D$^+$) $^{29}$Si NMR transition $\nu_{\rm Si0}$. Energy level diagram for the $^{29}$Si nuclear spin with and ionized $^{31}$P donor. The $\ket{\Uparrow}$ state is highest in energy as a result of the negative value of the $^{29}$Si nuclear gyromagnetic ratio. (\textbf{f}) NMR signal for the $\nu_{\rm Si0}$ transition, with a Lorentzian fit.}
\end{figure}

Next we perform an NMR experiment on the single $^{29}$Si nucleus. The whole system is described by the spin Hamiltonian~\cite{Feher1959I_PR, Levitt2008_Book}:
\begin{eqnarray}
\mathcal{H} = &B_0\left(\gamma_eS_z - \gamma_PI_z^{\rm P} - \gamma_{\rm Si}I_z^{\rm Si}\right) + A_P S\cdot I^{\rm P} \nonumber\\*
&+ A_{\rm Si}S\cdot I^{\rm Si} \label{eq:H29si}
\end{eqnarray}
where $S = I^{\rm P} = I^{\rm Si} = 1/2$ are the electron, $^{31}$P and $^{29}$Si spin operators and $\gamma_e = 28$~GHz/T, $\gamma_P = 17.23$~MHz/T, $\gamma_{\rm Si} = -8.458$~MHz/T (Ref.~\citenum{Hale1969I_PR}) are their respective gyromagnetic ratios. We assume that the electron-$^{29}$Si interaction $A_{\rm Si}$ is dominated by a contact hyperfine term, i.e. we omit the dipolar coupling between $^{29}$Si and the electron. This omission is justified by the fact that we observe an extremely small probability to flip the $^{29}$Si spin by ionizing/neutralizing the donor ($\sim 1$ flip every 100,000 readout events), which indicates that the secular approximation for the electron-nuclear interaction is almost exact, and non-diagonal interaction terms are negligible. For this reason, the nuclear spin measurement is almost exactly quantum-nondemolition (QND)~\cite{Braginsky1996_RMP}.

Calling $\nu_{\rm Si1}$ the $^{29}$Si NMR frequency for a $\ket{\downarrow}$ electron, and $\nu_{\rm Si2}$ for $\ket{\uparrow}$ (Fig.~\ref{fig:figure2}b), one has $\nu_{\rm Si1,2} = \gamma_{\rm Si}B_0 \mp A_{\rm Si}/2$. Since the $^{29}$Si hyperfine splitting observed in Fig.~\ref{fig:figure1}b is $\sim 2.2$~MHz at $B_0 = 1.77$~T, we extract $\nu_{\rm Si1} \approx 13.88$~MHz and $\nu_{\rm Si2} \approx 16.08$~MHz. We then perform a NMR experiment where we first initialize the electron spin, for example $\ket{\uparrow}$, and apply a long NMR pulse at a frequency $\nu_{\rm NMR}$ before attempting to adiabatically invert and read the electron spin. The electron spin-up fraction $f_{\uparrow}\left(\nu_{\rm e2L/R}\right)$ is then recorded, where $\nu_{\rm e2L}$ and $\nu_{\rm e2R}$ are the $^{29}$Si spin-dependent ESR transition frequencies defined as $\nu_{\rm e2L,R} = \gamma_eB_0 \mp A_{\rm Si}/2$.

\begin{figure}[t!]
\includegraphics[width=1\columnwidth]{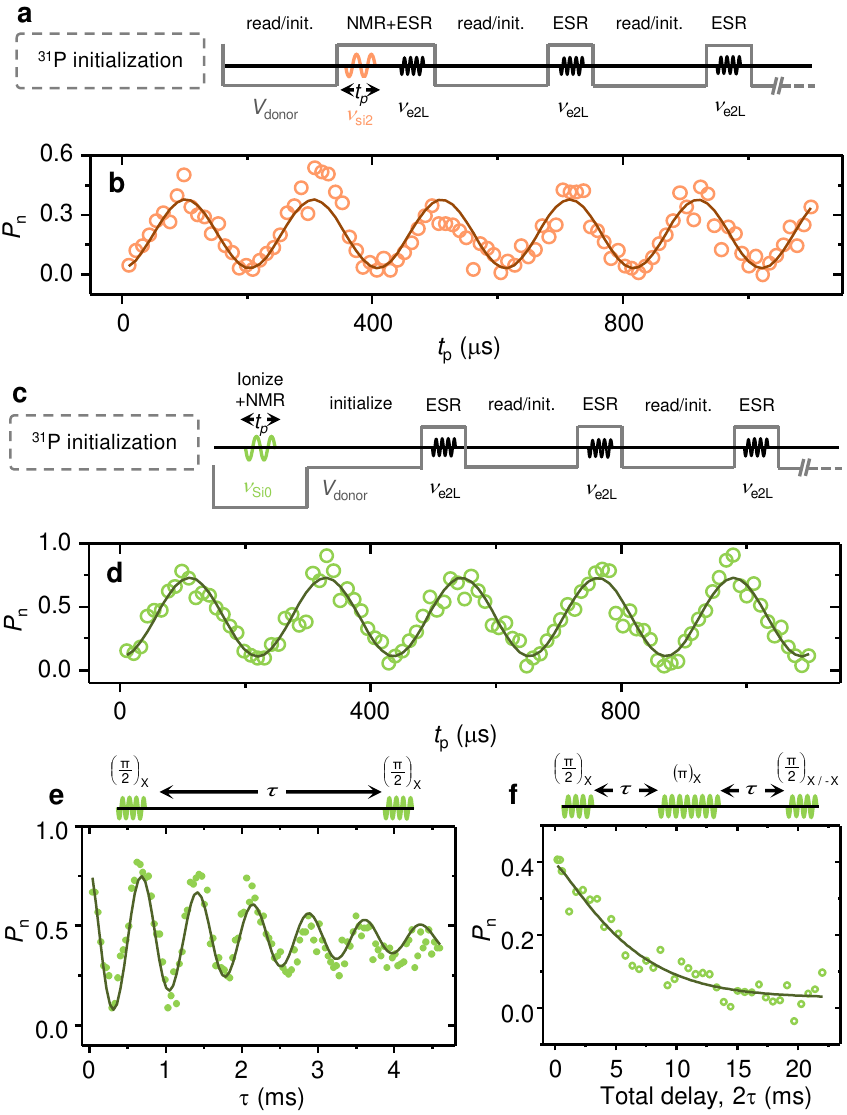}
\caption{\label{fig:figure3} (Color online) (\textbf{a}) Protocol and (\textbf{b}) measurement of single $^{29}$Si nuclear spin Rabi oscillations, i.e. nuclear spin flip probability $P_n$ as a function of the pulse duration $t_p$, for the neutral donor with $m_e = m_P = +1/2$. (\textbf{c}) Protocol and (\textbf{d}) measurement of $^{29}$Si Rabi oscillations with an ionized $^{31}$P donor. Fits for both curves (panels \textbf{b} and \textbf{d}) are of the form $P_n \propto \sin^2(\pi f_{\rm Rabi}t_p)$, where the Rabi frequency $f_{\rm rabi}$ is a free fitting parameter. (\textbf{e}) Ramsey fringe measurement. (\textbf{f}) Hahn echo decay measured with phase cycling (between X and $-$X) of the final $\pi/2$ pulse. Fits to data in panels \textbf{e} and \textbf{f} are described in the main text.}
\end{figure}

For each $\nu_{\rm NMR}$ we calculate $|\Delta f_{\uparrow}| = |f_{\uparrow}\left(\nu_{\rm e2R}\right) - f_{\uparrow}\left(\nu_{\rm e2L}\right)|$ and plot the result for the $\nu_{\rm Si2}$ transition in Fig.~\ref{fig:figure2}c. Off-resonance we find $|\Delta f_{\uparrow}| \approx 0.21$. At resonance, a randomization of the $^{29}$Si spin state produces an almost equal probability of having an ``active'' $\nu_{\rm e2L}$ or $\nu_{\rm e2R}$ transition. The trough observed at $\nu_{\rm NMR} = 16.11(2)$~MHz is remarkably close to the estimated value for $\nu_{\rm Si2}$.

The tunnel-coupled SET used for readout can also be utilized to ionize the $^{31}$P donor and perform NMR on the isolated $^{29}$Si nuclear spin (Fig.~\ref{fig:figure2}e). Here the NMR frequency is simply $\nu_{\rm Si0} = \gamma_{\rm Si}B_0$. The pulse sequence for such a measurement is shown in Fig.~\ref{fig:figure2}d with the resulting resonance plot in Fig.~\ref{fig:figure2}f. The trough at $\nu_{\rm NMR} = 14.99(2)$~MHz, together with the external magnetic field $B_0 = 1.77$~T -- calibrated using the measured $^{31}$P NMR frequencies~\cite{Pla2013_N} -- implies a gyromagnetic ratio of $|\gamma_{\rm Si}| = 8.47$~MHz/T, very close to the bulk value of 8.458~MHz/T (Ref.~\citenum{Hale1969I_PR}). These experiments also yield an accurate value for the hyperfine coupling $A_{\rm Si} = 2 \times (\nu_{\rm Si2} -\nu_{\rm Si0}) =  2.205(5)$~MHz.

We demonstrate the ability to coherently manipulate the $^{29}$Si nuclear spin -- with both a neutral (D$^0$) and ionized (D$^+$) donor -- by observing Rabi oscillations. The protocols for such measurements are illustrated in Figs.~\ref{fig:figure3}a and c, and the $^{29}$Si nuclear spin flip probabilities $P_n$ as a function of the pulse duration $t_p$ are shown in Figs.~\ref{fig:figure3}b and \ref{fig:figure3}d for the donor in the D$^0$ and D$^+$ charge states, respectively. The D$^+$ data displays higher visibility oscillations than the D$^0$ case, due to its immunity to electron spin state initialization errors.

Next we probe the coherence of the isolated (ionized donor) $^{29}$Si nuclear spin by performing Ramsey fringe and Hahn echo experiments (Fig.~\ref{fig:figure3}). Fitting the Ramsey data in Fig.~\ref{fig:figure3}e with a damped cosine function of the form $P_n = P_n(0)\cos\left(2\pi\Delta d\tau\right)\exp\left(-\tau/T_2^*\right)$ yields a dephasing time of $T_2^* = 2.4(3)$~ms. Also from this fit we get $\Delta d$, the average detuning from resonance, which enables us to provide a more accurate estimate of the gyromagnetic ratio $\gamma_{\rm Si} =  8.460(2)$~MHz/T. The echo decay curve of Fig.~\ref{fig:figure3}f, fitted with an exponential function $y = y(0)\exp\left((-2\tau/T_2)^b\right)$, reveals a coherence time of $T_2 = 6.3(7)$~ms and an exponent $b = 1.2(2)$. The coherence time is in excellent agreement with Hahn echo measurements in bulk~\cite{Dementyev2003_PRB}, where decoherence is caused by the dipole interactions with other $^{29}$Si nuclear spins.

\begin{figure}[t!]
\includegraphics[width=1\columnwidth]{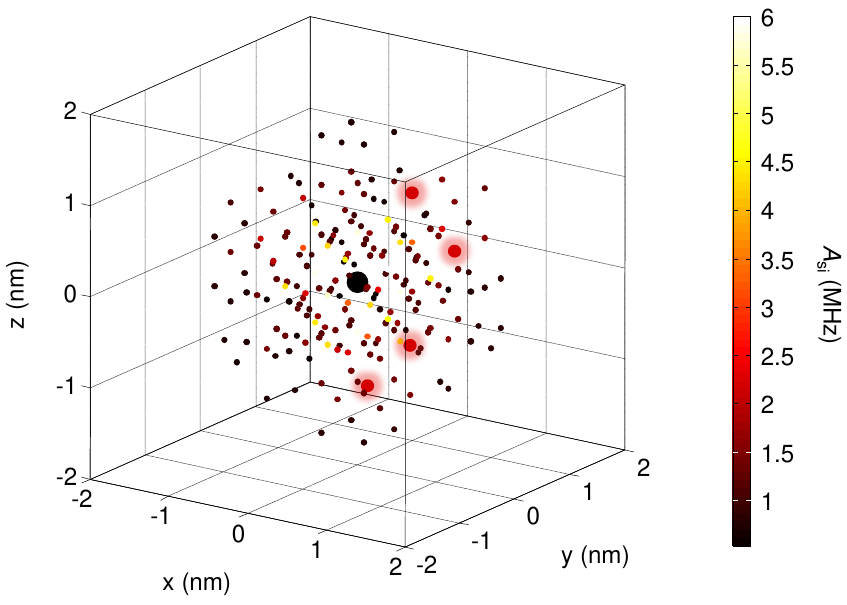}
\caption{\label{fig:figure4} (Color online) Atomistic modeling to match the experimental hyperfine coupling. Plotted are the $^{29}$Si nuclear spins with known hyperfine couplings~\cite{Hale1969I_PR}. The color-scale indicates the hyperfine interaction at each site, rescaled to reflect the distorted donor electron wavefunction in our specific device. The lattice sites that correspond to couplings within the range $2.15-2.25$~MHz are shown as larger circles.}
\end{figure}

The individual hyperfine couplings between $^{29}$Si nuclei and a donor-bound electron are known from early work in bulk samples~\cite{Hale1969I_PR,Hale1969II_PR,Hale1971_PRB,Ivey1972_PRL,Ivey1975I_PRB,Ivey1975II_PRB}. By adapting metrology techniques demonstrated for $^{31}$P~\cite{Mohiyaddin2013_NL}, we can narrow down the possible locations of the $^{29}$Si atom measured here. A device-specific wavefunction was obtained by first calculating, with a finite-elements Poisson equation solver, the electrostatic potential profile surrounding the donor, then solving the full atomistic tight-binding Hamiltonian with the tool Nano Electronic MOdeling 3D (NEMO 3D)~\cite{Klimeck2007_IEEE}. Calculating the shift from the bulk value in the probability density of the electron wavefunction $\left|\psi(r_0)\right|^2$ at each lattice site, allows us to appropriately scale the $^{29}$Si Fermi contact hyperfine splittings from bulk data. Figure~\ref{fig:figure4} shows a 3D plot of the $^{31}$P donor (large black circle) and the surrounding $^{29}$Si nuclei with known hyperfine constants. The $^{29}$Si nuclei with couplings in the range $2.15-2.25$~MHz are plot as enlarged circles. They all belong to a $(3,3,7)$ shell at 1.11~nm distance from the $^{31}$P nucleus \cite{supp}. We have thus been able to narrow down the location of our $^{29}$Si atom to 4 out of a known $\sim 150$ possible sites.

In conclusion, we have performed electrical single-shot QND readout on a single $^{29}$Si nuclear spin, and demonstrated its coherent control though Rabi, Ramsey and Hahn echo experiments, which yield coherence values similar to those observed in bulk samples. While the isotopic purification of $^{28}$Si is an exciting avenue to achieve the best possible coherence and fidelity benchmarks \cite{Muhonen2014}, the present work shows that isolated $^{29}$Si nuclear spins can be utilized as an additional resource~\cite{Robledo2011_N} for quantum information processing in silicon.

We thank R.P. Starrett, D. Barber, C.Y. Yang and R. Szymanski for technical assistance, and W.A. Coish for discussions. This research was funded by the Australian Research Council Centre of Excellence for Quantum Computation and Communication Technology (project number CE110001027) and the US Army Research Office (W911NF-13-1-0024). We acknowledge support from the Australian National Fabrication Facility. Computational resources on nanoHUB.org, funded by the NSF grant EEC-0228390, were used in this work.


\bibliography{29Si}

\begin{thebibliography}{39}%
\makeatletter
\providecommand \@ifxundefined [1]{%
 \@ifx{#1\undefined}
}%
\providecommand \@ifnum [1]{%
 \ifnum #1\expandafter \@firstoftwo
 \else \expandafter \@secondoftwo
 \fi
}%
\providecommand \@ifx [1]{%
 \ifx #1\expandafter \@firstoftwo
 \else \expandafter \@secondoftwo
 \fi
}%
\providecommand \natexlab [1]{#1}%
\providecommand \enquote  [1]{``#1''}%
\providecommand \bibnamefont  [1]{#1}%
\providecommand \bibfnamefont [1]{#1}%
\providecommand \citenamefont [1]{#1}%
\providecommand \href@noop [0]{\@secondoftwo}%
\providecommand \href [0]{\begingroup \@sanitize@url \@href}%
\providecommand \@href[1]{\@@startlink{#1}\@@href}%
\providecommand \@@href[1]{\endgroup#1\@@endlink}%
\providecommand \@sanitize@url [0]{\catcode `\\12\catcode `\$12\catcode
  `\&12\catcode `\#12\catcode `\^12\catcode `\_12\catcode `\%12\relax}%
\providecommand \@@startlink[1]{}%
\providecommand \@@endlink[0]{}%
\providecommand \url  [0]{\begingroup\@sanitize@url \@url }%
\providecommand \@url [1]{\endgroup\@href {#1}{\urlprefix }}%
\providecommand \urlprefix  [0]{URL }%
\providecommand \Eprint [0]{\href }%
\providecommand \doibase [0]{http://dx.doi.org/}%
\providecommand \selectlanguage [0]{\@gobble}%
\providecommand \bibinfo  [0]{\@secondoftwo}%
\providecommand \bibfield  [0]{\@secondoftwo}%
\providecommand \translation [1]{[#1]}%
\providecommand \BibitemOpen [0]{}%
\providecommand \bibitemStop [0]{}%
\providecommand \bibitemNoStop [0]{.\EOS\space}%
\providecommand \EOS [0]{\spacefactor3000\relax}%
\providecommand \BibitemShut  [1]{\csname bibitem#1\endcsname}%
\let\auto@bib@innerbib\@empty
\bibitem [{\citenamefont {Klauder}\ and\ \citenamefont
  {Anderson}(1962)}]{Klauder1962_PR}%
  \BibitemOpen
  \bibfield  {author} {\bibinfo {author} {\bibfnamefont {J.~R.}\ \bibnamefont
  {Klauder}}\ and\ \bibinfo {author} {\bibfnamefont {P.~W.}\ \bibnamefont
  {Anderson}},\ }\href {\doibase 10.1103/PhysRev.125.912} {\bibfield  {journal}
  {\bibinfo  {journal} {Phys. Rev.}\ }\textbf {\bibinfo {volume} {125}},\
  \bibinfo {pages} {912} (\bibinfo {year} {1962})}\BibitemShut {NoStop}%
\bibitem [{\citenamefont {Pla}\ \emph {et~al.}(2012)\citenamefont {Pla},
  \citenamefont {Tan}, \citenamefont {Dehollain}, \citenamefont {Lim},
  \citenamefont {Morton}, \citenamefont {Jamieson}, \citenamefont {Dzurak},\
  and\ \citenamefont {Morello}}]{Pla2012_N}%
  \BibitemOpen
  \bibfield  {author} {\bibinfo {author} {\bibfnamefont {J.~J.}\ \bibnamefont
  {Pla}}, \bibinfo {author} {\bibfnamefont {K.~Y.}\ \bibnamefont {Tan}},
  \bibinfo {author} {\bibfnamefont {J.~P.}\ \bibnamefont {Dehollain}}, \bibinfo
  {author} {\bibfnamefont {W.~H.}\ \bibnamefont {Lim}}, \bibinfo {author}
  {\bibfnamefont {J.~J.}\ \bibnamefont {Morton}}, \bibinfo {author}
  {\bibfnamefont {D.~N.}\ \bibnamefont {Jamieson}}, \bibinfo {author}
  {\bibfnamefont {A.~S.}\ \bibnamefont {Dzurak}}, \ and\ \bibinfo {author}
  {\bibfnamefont {A.}~\bibnamefont {Morello}},\ }\href@noop {} {\bibfield
  {journal} {\bibinfo  {journal} {Nature}\ }\textbf {\bibinfo {volume} {489}},\
  \bibinfo {pages} {541} (\bibinfo {year} {2012})}\BibitemShut {NoStop}%
\bibitem [{\citenamefont {Pla}\ \emph {et~al.}(2013)\citenamefont {Pla},
  \citenamefont {Tan}, \citenamefont {Dehollain}, \citenamefont {Lim},
  \citenamefont {Morton}, \citenamefont {Zwanenburg}, \citenamefont {Jamieson},
  \citenamefont {Dzurak},\ and\ \citenamefont {Morello}}]{Pla2013_N}%
  \BibitemOpen
  \bibfield  {author} {\bibinfo {author} {\bibfnamefont {J.~J.}\ \bibnamefont
  {Pla}}, \bibinfo {author} {\bibfnamefont {K.~Y.}\ \bibnamefont {Tan}},
  \bibinfo {author} {\bibfnamefont {J.~P.}\ \bibnamefont {Dehollain}}, \bibinfo
  {author} {\bibfnamefont {W.~H.}\ \bibnamefont {Lim}}, \bibinfo {author}
  {\bibfnamefont {J.~J.}\ \bibnamefont {Morton}}, \bibinfo {author}
  {\bibfnamefont {F.~A.}\ \bibnamefont {Zwanenburg}}, \bibinfo {author}
  {\bibfnamefont {D.~N.}\ \bibnamefont {Jamieson}}, \bibinfo {author}
  {\bibfnamefont {A.~S.}\ \bibnamefont {Dzurak}}, \ and\ \bibinfo {author}
  {\bibfnamefont {A.}~\bibnamefont {Morello}},\ }\href@noop {} {\bibfield
  {journal} {\bibinfo  {journal} {Nature}\ }\textbf {\bibinfo {volume} {496}},\
  \bibinfo {pages} {334} (\bibinfo {year} {2013})}\BibitemShut {NoStop}%
\bibitem [{\citenamefont {Hanson}\ \emph {et~al.}(2008)\citenamefont {Hanson},
  \citenamefont {Dobrovitski}, \citenamefont {Feiguin}, \citenamefont {Gywat},\
  and\ \citenamefont {Awschalom}}]{Hanson2008_S}%
  \BibitemOpen
  \bibfield  {author} {\bibinfo {author} {\bibfnamefont {R.}~\bibnamefont
  {Hanson}}, \bibinfo {author} {\bibfnamefont {V.}~\bibnamefont {Dobrovitski}},
  \bibinfo {author} {\bibfnamefont {A.}~\bibnamefont {Feiguin}}, \bibinfo
  {author} {\bibfnamefont {O.}~\bibnamefont {Gywat}}, \ and\ \bibinfo {author}
  {\bibfnamefont {D.}~\bibnamefont {Awschalom}},\ }\href@noop {} {\bibfield
  {journal} {\bibinfo  {journal} {Science}\ }\textbf {\bibinfo {volume}
  {320}},\ \bibinfo {pages} {352} (\bibinfo {year} {2008})}\BibitemShut
  {NoStop}%
\bibitem [{\citenamefont {Coish}\ and\ \citenamefont
  {Loss}(2005)}]{Coish2005_PRB}%
  \BibitemOpen
  \bibfield  {author} {\bibinfo {author} {\bibfnamefont {W.~A.}\ \bibnamefont
  {Coish}}\ and\ \bibinfo {author} {\bibfnamefont {D.}~\bibnamefont {Loss}},\
  }\href {\doibase 10.1103/PhysRevB.72.125337} {\bibfield  {journal} {\bibinfo
  {journal} {Phys. Rev. B}\ }\textbf {\bibinfo {volume} {72}},\ \bibinfo
  {pages} {125337} (\bibinfo {year} {2005})}\BibitemShut {NoStop}%
\bibitem [{\citenamefont {Yao}\ \emph {et~al.}(2006)\citenamefont {Yao},
  \citenamefont {Liu},\ and\ \citenamefont {Sham}}]{Yao2006_PRB}%
  \BibitemOpen
  \bibfield  {author} {\bibinfo {author} {\bibfnamefont {W.}~\bibnamefont
  {Yao}}, \bibinfo {author} {\bibfnamefont {R.-B.}\ \bibnamefont {Liu}}, \ and\
  \bibinfo {author} {\bibfnamefont {L.}~\bibnamefont {Sham}},\ }\href@noop {}
  {\bibfield  {journal} {\bibinfo  {journal} {Physical Review B}\ }\textbf
  {\bibinfo {volume} {74}},\ \bibinfo {pages} {195301} (\bibinfo {year}
  {2006})}\BibitemShut {NoStop}%
\bibitem [{\citenamefont {Dutt}\ \emph {et~al.}(2007)\citenamefont {Dutt},
  \citenamefont {Childress}, \citenamefont {Jiang}, \citenamefont {Togan},
  \citenamefont {Maze}, \citenamefont {Jelezko}, \citenamefont {Zibrov},
  \citenamefont {Hemmer},\ and\ \citenamefont {Lukin}}]{Dutt2007_S}%
  \BibitemOpen
  \bibfield  {author} {\bibinfo {author} {\bibfnamefont {M.~V.~G.}\
  \bibnamefont {Dutt}}, \bibinfo {author} {\bibfnamefont {L.}~\bibnamefont
  {Childress}}, \bibinfo {author} {\bibfnamefont {L.}~\bibnamefont {Jiang}},
  \bibinfo {author} {\bibfnamefont {E.}~\bibnamefont {Togan}}, \bibinfo
  {author} {\bibfnamefont {J.}~\bibnamefont {Maze}}, \bibinfo {author}
  {\bibfnamefont {F.}~\bibnamefont {Jelezko}}, \bibinfo {author} {\bibfnamefont
  {A.~S.}\ \bibnamefont {Zibrov}}, \bibinfo {author} {\bibfnamefont {P.~R.}\
  \bibnamefont {Hemmer}}, \ and\ \bibinfo {author} {\bibfnamefont {M.~D.}\
  \bibnamefont {Lukin}},\ }\href {\doibase 10.1126/science.1139831} {\bibfield
  {journal} {\bibinfo  {journal} {Science}\ }\textbf {\bibinfo {volume}
  {316}},\ \bibinfo {pages} {1312} (\bibinfo {year} {2007})}\BibitemShut
  {NoStop}%
\bibitem [{\citenamefont {Robledo}\ \emph {et~al.}(2011)\citenamefont
  {Robledo}, \citenamefont {Childress}, \citenamefont {Bernien}, \citenamefont
  {Hensen}, \citenamefont {Alkemade},\ and\ \citenamefont
  {Hanson}}]{Robledo2011_N}%
  \BibitemOpen
  \bibfield  {author} {\bibinfo {author} {\bibfnamefont {L.}~\bibnamefont
  {Robledo}}, \bibinfo {author} {\bibfnamefont {L.}~\bibnamefont {Childress}},
  \bibinfo {author} {\bibfnamefont {H.}~\bibnamefont {Bernien}}, \bibinfo
  {author} {\bibfnamefont {B.}~\bibnamefont {Hensen}}, \bibinfo {author}
  {\bibfnamefont {P.~F.}\ \bibnamefont {Alkemade}}, \ and\ \bibinfo {author}
  {\bibfnamefont {R.}~\bibnamefont {Hanson}},\ }\href@noop {} {\bibfield
  {journal} {\bibinfo  {journal} {Nature}\ }\textbf {\bibinfo {volume} {477}},\
  \bibinfo {pages} {574} (\bibinfo {year} {2011})}\BibitemShut {NoStop}%
\bibitem [{\citenamefont {Waldherr}\ \emph {et~al.}(2014)\citenamefont
  {Waldherr}, \citenamefont {Wang}, \citenamefont {Zaiser}, \citenamefont
  {Jamali}, \citenamefont {Schulte-Herbr{\"u}ggen}, \citenamefont {Abe},
  \citenamefont {Ohshima}, \citenamefont {Isoya}, \citenamefont {Du},
  \citenamefont {Neumann} \emph {et~al.}}]{Waldherr2014_N}%
  \BibitemOpen
  \bibfield  {author} {\bibinfo {author} {\bibfnamefont {G.}~\bibnamefont
  {Waldherr}}, \bibinfo {author} {\bibfnamefont {Y.}~\bibnamefont {Wang}},
  \bibinfo {author} {\bibfnamefont {S.}~\bibnamefont {Zaiser}}, \bibinfo
  {author} {\bibfnamefont {M.}~\bibnamefont {Jamali}}, \bibinfo {author}
  {\bibfnamefont {T.}~\bibnamefont {Schulte-Herbr{\"u}ggen}}, \bibinfo {author}
  {\bibfnamefont {H.}~\bibnamefont {Abe}}, \bibinfo {author} {\bibfnamefont
  {T.}~\bibnamefont {Ohshima}}, \bibinfo {author} {\bibfnamefont
  {J.}~\bibnamefont {Isoya}}, \bibinfo {author} {\bibfnamefont
  {J.}~\bibnamefont {Du}}, \bibinfo {author} {\bibfnamefont {P.}~\bibnamefont
  {Neumann}},  \emph {et~al.},\ }\href {\doibase 10.1038/nature12919}
  {\bibfield  {journal} {\bibinfo  {journal} {Nature}\ }\textbf {\bibinfo
  {volume} {506}},\ \bibinfo {pages} {204} (\bibinfo {year}
  {2014})}\BibitemShut {NoStop}%
\bibitem [{\citenamefont {Taminiau}\ \emph {et~al.}(2014)\citenamefont
  {Taminiau}, \citenamefont {Cramer}, \citenamefont {van~der Sar},
  \citenamefont {Dobrovitski},\ and\ \citenamefont {Hanson}}]{Taminiau2014_N}%
  \BibitemOpen
  \bibfield  {author} {\bibinfo {author} {\bibfnamefont {T.~H.}\ \bibnamefont
  {Taminiau}}, \bibinfo {author} {\bibfnamefont {J.}~\bibnamefont {Cramer}},
  \bibinfo {author} {\bibfnamefont {T.}~\bibnamefont {van~der Sar}}, \bibinfo
  {author} {\bibfnamefont {V.~V.}\ \bibnamefont {Dobrovitski}}, \ and\ \bibinfo
  {author} {\bibfnamefont {R.}~\bibnamefont {Hanson}},\ }\href@noop {}
  {\bibfield  {journal} {\bibinfo  {journal} {Nature nanotechnology}\ }\textbf
  {\bibinfo {volume} {9}},\ \bibinfo {pages} {171} (\bibinfo {year}
  {2014})}\BibitemShut {NoStop}%
\bibitem [{\citenamefont {Nickerson}\ \emph {et~al.}(2013)\citenamefont
  {Nickerson}, \citenamefont {Li},\ and\ \citenamefont
  {Benjamin}}]{Nickerson2013_NC}%
  \BibitemOpen
  \bibfield  {author} {\bibinfo {author} {\bibfnamefont {N.~H.}\ \bibnamefont
  {Nickerson}}, \bibinfo {author} {\bibfnamefont {Y.}~\bibnamefont {Li}}, \
  and\ \bibinfo {author} {\bibfnamefont {S.~C.}\ \bibnamefont {Benjamin}},\
  }\href@noop {} {\bibfield  {journal} {\bibinfo  {journal} {Nature
  communications}\ }\textbf {\bibinfo {volume} {4}},\ \bibinfo {pages} {1756}
  (\bibinfo {year} {2013})}\BibitemShut {NoStop}%
\bibitem [{\citenamefont {Ladd}\ \emph {et~al.}(2002)\citenamefont {Ladd},
  \citenamefont {Goldman}, \citenamefont {Yamaguchi}, \citenamefont {Yamamoto},
  \citenamefont {Abe},\ and\ \citenamefont {Itoh}}]{Ladd2002_PRL}%
  \BibitemOpen
  \bibfield  {author} {\bibinfo {author} {\bibfnamefont {T.~D.}\ \bibnamefont
  {Ladd}}, \bibinfo {author} {\bibfnamefont {J.~R.}\ \bibnamefont {Goldman}},
  \bibinfo {author} {\bibfnamefont {F.}~\bibnamefont {Yamaguchi}}, \bibinfo
  {author} {\bibfnamefont {Y.}~\bibnamefont {Yamamoto}}, \bibinfo {author}
  {\bibfnamefont {E.}~\bibnamefont {Abe}}, \ and\ \bibinfo {author}
  {\bibfnamefont {K.~M.}\ \bibnamefont {Itoh}},\ }\href {\doibase
  10.1103/PhysRevLett.89.017901} {\bibfield  {journal} {\bibinfo  {journal}
  {Phys. Rev. Lett.}\ }\textbf {\bibinfo {volume} {89}},\ \bibinfo {pages}
  {017901} (\bibinfo {year} {2002})}\BibitemShut {NoStop}%
\bibitem [{\citenamefont {Itoh}(2005)}]{Itoh2005_SSC}%
  \BibitemOpen
  \bibfield  {author} {\bibinfo {author} {\bibfnamefont {K.~M.}\ \bibnamefont
  {Itoh}},\ }\href {\doibase 10.1016/j.ssc.2004.12.045} {\bibfield  {journal}
  {\bibinfo  {journal} {Solid State Communications}\ }\textbf {\bibinfo
  {volume} {133}},\ \bibinfo {pages} {747 } (\bibinfo {year}
  {2005})}\BibitemShut {NoStop}%
\bibitem [{\citenamefont {Morello}\ \emph {et~al.}(2009)\citenamefont
  {Morello}, \citenamefont {Escott}, \citenamefont {Huebl}, \citenamefont
  {Willems~van Beveren}, \citenamefont {Hollenberg}, \citenamefont {Jamieson},
  \citenamefont {Dzurak},\ and\ \citenamefont {Clark}}]{Morello2009_PRB}%
  \BibitemOpen
  \bibfield  {author} {\bibinfo {author} {\bibfnamefont {A.}~\bibnamefont
  {Morello}}, \bibinfo {author} {\bibfnamefont {C.~C.}\ \bibnamefont {Escott}},
  \bibinfo {author} {\bibfnamefont {H.}~\bibnamefont {Huebl}}, \bibinfo
  {author} {\bibfnamefont {L.~H.}\ \bibnamefont {Willems~van Beveren}},
  \bibinfo {author} {\bibfnamefont {L.~C.~L.}\ \bibnamefont {Hollenberg}},
  \bibinfo {author} {\bibfnamefont {D.~N.}\ \bibnamefont {Jamieson}}, \bibinfo
  {author} {\bibfnamefont {A.~S.}\ \bibnamefont {Dzurak}}, \ and\ \bibinfo
  {author} {\bibfnamefont {R.~G.}\ \bibnamefont {Clark}},\ }\href {\doibase
  10.1103/PhysRevB.80.081307} {\bibfield  {journal} {\bibinfo  {journal} {Phys.
  Rev. B}\ }\textbf {\bibinfo {volume} {80}},\ \bibinfo {pages} {081307}
  (\bibinfo {year} {2009})}\BibitemShut {NoStop}%
\bibitem [{\citenamefont {Jamieson}\ \emph {et~al.}(2005)\citenamefont
  {Jamieson}, \citenamefont {Yang}, \citenamefont {Hopf}, \citenamefont
  {Hearne}, \citenamefont {Pakes}, \citenamefont {Prawer}, \citenamefont
  {Mitic}, \citenamefont {Gauja}, \citenamefont {Andresen}, \citenamefont
  {Hudson}, \citenamefont {Dzurak},\ and\ \citenamefont
  {Clark}}]{Jamieson2005_APL}%
  \BibitemOpen
  \bibfield  {author} {\bibinfo {author} {\bibfnamefont {D.~N.}\ \bibnamefont
  {Jamieson}}, \bibinfo {author} {\bibfnamefont {C.}~\bibnamefont {Yang}},
  \bibinfo {author} {\bibfnamefont {T.}~\bibnamefont {Hopf}}, \bibinfo {author}
  {\bibfnamefont {S.~M.}\ \bibnamefont {Hearne}}, \bibinfo {author}
  {\bibfnamefont {C.~I.}\ \bibnamefont {Pakes}}, \bibinfo {author}
  {\bibfnamefont {S.}~\bibnamefont {Prawer}}, \bibinfo {author} {\bibfnamefont
  {M.}~\bibnamefont {Mitic}}, \bibinfo {author} {\bibfnamefont
  {E.}~\bibnamefont {Gauja}}, \bibinfo {author} {\bibfnamefont {S.~E.}\
  \bibnamefont {Andresen}}, \bibinfo {author} {\bibfnamefont {F.~E.}\
  \bibnamefont {Hudson}}, \bibinfo {author} {\bibfnamefont {A.~S.}\
  \bibnamefont {Dzurak}}, \ and\ \bibinfo {author} {\bibfnamefont {R.~G.}\
  \bibnamefont {Clark}},\ }\href {\doibase 10.1063/1.1925320} {\bibfield
  {journal} {\bibinfo  {journal} {Applied Physics Letters}\ }\textbf {\bibinfo
  {volume} {86}},\ \bibinfo {eid} {202101} (\bibinfo {year}
  {2005})}\BibitemShut {NoStop}%
\bibitem [{\citenamefont {Angus}\ \emph {et~al.}(2007)\citenamefont {Angus},
  \citenamefont {Ferguson}, \citenamefont {Dzurak},\ and\ \citenamefont
  {Clark}}]{Angus2007_NL}%
  \BibitemOpen
  \bibfield  {author} {\bibinfo {author} {\bibfnamefont {S.~J.}\ \bibnamefont
  {Angus}}, \bibinfo {author} {\bibfnamefont {A.~J.}\ \bibnamefont {Ferguson}},
  \bibinfo {author} {\bibfnamefont {A.~S.}\ \bibnamefont {Dzurak}}, \ and\
  \bibinfo {author} {\bibfnamefont {R.~G.}\ \bibnamefont {Clark}},\ }\href
  {\doibase 10.1021/nl070949k} {\bibfield  {journal} {\bibinfo  {journal} {Nano
  Letters}\ }\textbf {\bibinfo {volume} {7}},\ \bibinfo {pages} {2051}
  (\bibinfo {year} {2007})}\BibitemShut {NoStop}%
\bibitem [{\citenamefont {Dehollain}\ \emph {et~al.}(2013)\citenamefont
  {Dehollain}, \citenamefont {Pla}, \citenamefont {Siew}, \citenamefont {Tan},
  \citenamefont {Dzurak},\ and\ \citenamefont {Morello}}]{Dehollain2013_Na}%
  \BibitemOpen
  \bibfield  {author} {\bibinfo {author} {\bibfnamefont {J.~P.}\ \bibnamefont
  {Dehollain}}, \bibinfo {author} {\bibfnamefont {J.~J.}\ \bibnamefont {Pla}},
  \bibinfo {author} {\bibfnamefont {E.}~\bibnamefont {Siew}}, \bibinfo {author}
  {\bibfnamefont {K.~Y.}\ \bibnamefont {Tan}}, \bibinfo {author} {\bibfnamefont
  {A.~S.}\ \bibnamefont {Dzurak}}, \ and\ \bibinfo {author} {\bibfnamefont
  {A.}~\bibnamefont {Morello}},\ }\href {\doibase
  10.1088/0957-4484/24/1/015202} {\bibfield  {journal} {\bibinfo  {journal}
  {Nanotechnology}\ }\textbf {\bibinfo {volume} {24}},\ \bibinfo {pages}
  {015202} (\bibinfo {year} {2013})}\BibitemShut {NoStop}%
\bibitem [{\citenamefont {Tyryshkin}\ \emph {et~al.}(2003)\citenamefont
  {Tyryshkin}, \citenamefont {Lyon}, \citenamefont {Astashkin},\ and\
  \citenamefont {Raitsimring}}]{Tyryshkin2003_PRB}%
  \BibitemOpen
  \bibfield  {author} {\bibinfo {author} {\bibfnamefont {A.~M.}\ \bibnamefont
  {Tyryshkin}}, \bibinfo {author} {\bibfnamefont {S.~A.}\ \bibnamefont {Lyon}},
  \bibinfo {author} {\bibfnamefont {A.~V.}\ \bibnamefont {Astashkin}}, \ and\
  \bibinfo {author} {\bibfnamefont {A.~M.}\ \bibnamefont {Raitsimring}},\
  }\href {\doibase 10.1103/PhysRevB.68.193207} {\bibfield  {journal} {\bibinfo
  {journal} {Phys. Rev. B}\ }\textbf {\bibinfo {volume} {68}},\ \bibinfo
  {pages} {193207} (\bibinfo {year} {2003})}\BibitemShut {NoStop}%
\bibitem [{\citenamefont {Hale}\ and\ \citenamefont
  {Mieher}(1969{\natexlab{a}})}]{Hale1969I_PR}%
  \BibitemOpen
  \bibfield  {author} {\bibinfo {author} {\bibfnamefont {E.~B.}\ \bibnamefont
  {Hale}}\ and\ \bibinfo {author} {\bibfnamefont {R.~L.}\ \bibnamefont
  {Mieher}},\ }\href {\doibase 10.1103/PhysRev.184.739} {\bibfield  {journal}
  {\bibinfo  {journal} {Phys. Rev.}\ }\textbf {\bibinfo {volume} {184}},\
  \bibinfo {pages} {739} (\bibinfo {year} {1969}{\natexlab{a}})}\BibitemShut
  {NoStop}%
\bibitem [{\citenamefont {Laird}\ \emph {et~al.}(2007)\citenamefont {Laird},
  \citenamefont {Barthel}, \citenamefont {Rashba}, \citenamefont {Marcus},
  \citenamefont {Hanson},\ and\ \citenamefont {Gossard}}]{Laird2007_PRL}%
  \BibitemOpen
  \bibfield  {author} {\bibinfo {author} {\bibfnamefont {E.}~\bibnamefont
  {Laird}}, \bibinfo {author} {\bibfnamefont {C.}~\bibnamefont {Barthel}},
  \bibinfo {author} {\bibfnamefont {E.}~\bibnamefont {Rashba}}, \bibinfo
  {author} {\bibfnamefont {C.}~\bibnamefont {Marcus}}, \bibinfo {author}
  {\bibfnamefont {M.}~\bibnamefont {Hanson}}, \ and\ \bibinfo {author}
  {\bibfnamefont {A.}~\bibnamefont {Gossard}},\ }\href@noop {} {\bibfield
  {journal} {\bibinfo  {journal} {Physical review letters}\ }\textbf {\bibinfo
  {volume} {99}},\ \bibinfo {pages} {246601} (\bibinfo {year}
  {2007})}\BibitemShut {NoStop}%
\bibitem [{\citenamefont {Rudner}\ and\ \citenamefont
  {Levitov}(2007)}]{Rudner2007_PRL}%
  \BibitemOpen
  \bibfield  {author} {\bibinfo {author} {\bibfnamefont {M.}~\bibnamefont
  {Rudner}}\ and\ \bibinfo {author} {\bibfnamefont {L.}~\bibnamefont
  {Levitov}},\ }\href@noop {} {\bibfield  {journal} {\bibinfo  {journal}
  {Physical review letters}\ }\textbf {\bibinfo {volume} {99}},\ \bibinfo
  {pages} {036602} (\bibinfo {year} {2007})}\BibitemShut {NoStop}%
\bibitem [{\citenamefont {Rudner}\ \emph {et~al.}(2011)\citenamefont {Rudner},
  \citenamefont {Koppens}, \citenamefont {Folk}, \citenamefont {Vandersypen},\
  and\ \citenamefont {Levitov}}]{Rudner2011_PRB}%
  \BibitemOpen
  \bibfield  {author} {\bibinfo {author} {\bibfnamefont {M.}~\bibnamefont
  {Rudner}}, \bibinfo {author} {\bibfnamefont {F.}~\bibnamefont {Koppens}},
  \bibinfo {author} {\bibfnamefont {J.}~\bibnamefont {Folk}}, \bibinfo {author}
  {\bibfnamefont {L.}~\bibnamefont {Vandersypen}}, \ and\ \bibinfo {author}
  {\bibfnamefont {L.}~\bibnamefont {Levitov}},\ }\href@noop {} {\bibfield
  {journal} {\bibinfo  {journal} {Physical Review B}\ }\textbf {\bibinfo
  {volume} {84}},\ \bibinfo {pages} {075339} (\bibinfo {year}
  {2011})}\BibitemShut {NoStop}%
\bibitem [{\citenamefont {Urbaszek}\ \emph {et~al.}(2013)\citenamefont
  {Urbaszek}, \citenamefont {Marie}, \citenamefont {Amand}, \citenamefont
  {Krebs}, \citenamefont {Voisin}, \citenamefont {Maletinsky}, \citenamefont
  {H{\"o}gele},\ and\ \citenamefont {Imamoglu}}]{Urbaszek2013_RMP}%
  \BibitemOpen
  \bibfield  {author} {\bibinfo {author} {\bibfnamefont {B.}~\bibnamefont
  {Urbaszek}}, \bibinfo {author} {\bibfnamefont {X.}~\bibnamefont {Marie}},
  \bibinfo {author} {\bibfnamefont {T.}~\bibnamefont {Amand}}, \bibinfo
  {author} {\bibfnamefont {O.}~\bibnamefont {Krebs}}, \bibinfo {author}
  {\bibfnamefont {P.}~\bibnamefont {Voisin}}, \bibinfo {author} {\bibfnamefont
  {P.}~\bibnamefont {Maletinsky}}, \bibinfo {author} {\bibfnamefont
  {A.}~\bibnamefont {H{\"o}gele}}, \ and\ \bibinfo {author} {\bibfnamefont
  {A.}~\bibnamefont {Imamoglu}},\ }\href@noop {} {\bibfield  {journal}
  {\bibinfo  {journal} {Reviews of Modern Physics}\ }\textbf {\bibinfo {volume}
  {85}},\ \bibinfo {pages} {79} (\bibinfo {year} {2013})}\BibitemShut {NoStop}%
\bibitem [{\citenamefont {Yang}\ and\ \citenamefont
  {Sham}(2013)}]{Yang2013_PRB}%
  \BibitemOpen
  \bibfield  {author} {\bibinfo {author} {\bibfnamefont {W.}~\bibnamefont
  {Yang}}\ and\ \bibinfo {author} {\bibfnamefont {L.}~\bibnamefont {Sham}},\
  }\href@noop {} {\bibfield  {journal} {\bibinfo  {journal} {Physical Review
  B}\ }\textbf {\bibinfo {volume} {88}},\ \bibinfo {pages} {235304} (\bibinfo
  {year} {2013})}\BibitemShut {NoStop}%
\bibitem [{\citenamefont {Cappellaro}(2012)}]{Cappellaro2012_PRA}%
  \BibitemOpen
  \bibfield  {author} {\bibinfo {author} {\bibfnamefont {P.}~\bibnamefont
  {Cappellaro}},\ }\href {\doibase 10.1103/PhysRevA.85.030301} {\bibfield
  {journal} {\bibinfo  {journal} {Phys. Rev. A}\ }\textbf {\bibinfo {volume}
  {85}},\ \bibinfo {pages} {030301} (\bibinfo {year} {2012})}\BibitemShut
  {NoStop}%
\bibitem [{\citenamefont {Laucht}\ \emph {et~al.}(2014)\citenamefont {Laucht},
  \citenamefont {Kalra}, \citenamefont {Muhonen}, \citenamefont {Dehollain},
  \citenamefont {Mohiyaddin}, \citenamefont {Hudson}, \citenamefont {McCallum},
  \citenamefont {Jamieson}, \citenamefont {Dzurak},\ and\ \citenamefont
  {Morello}}]{Laucht2014_APL}%
  \BibitemOpen
  \bibfield  {author} {\bibinfo {author} {\bibfnamefont {A.}~\bibnamefont
  {Laucht}}, \bibinfo {author} {\bibfnamefont {R.}~\bibnamefont {Kalra}},
  \bibinfo {author} {\bibfnamefont {J.~T.}\ \bibnamefont {Muhonen}}, \bibinfo
  {author} {\bibfnamefont {J.~P.}\ \bibnamefont {Dehollain}}, \bibinfo {author}
  {\bibfnamefont {F.~A.}\ \bibnamefont {Mohiyaddin}}, \bibinfo {author}
  {\bibfnamefont {F.}~\bibnamefont {Hudson}}, \bibinfo {author} {\bibfnamefont
  {J.~C.}\ \bibnamefont {McCallum}}, \bibinfo {author} {\bibfnamefont {D.~N.}\
  \bibnamefont {Jamieson}}, \bibinfo {author} {\bibfnamefont {A.~S.}\
  \bibnamefont {Dzurak}}, \ and\ \bibinfo {author} {\bibfnamefont
  {A.}~\bibnamefont {Morello}},\ }\href
  {http://scitation.aip.org/content/aip/journal/apl/104/9/10.1063/1.4867905}
  {\bibfield  {journal} {\bibinfo  {journal} {Applied Physics Letters}\
  }\textbf {\bibinfo {volume} {104}},\ \bibinfo {eid} {092115} (\bibinfo {year}
  {2014})}\BibitemShut {NoStop}%
\bibitem [{sup()}]{supp}%
  \BibitemOpen
  \href@noop {} {\ }\bibinfo {note} {See Supplemental Material for extended
  figure captions}\BibitemShut {NoStop}%
\bibitem [{\citenamefont {Feher}(1959)}]{Feher1959I_PR}%
  \BibitemOpen
  \bibfield  {author} {\bibinfo {author} {\bibfnamefont {G.}~\bibnamefont
  {Feher}},\ }\href {\doibase 10.1103/PhysRev.114.1219} {\bibfield  {journal}
  {\bibinfo  {journal} {Phys. Rev.}\ }\textbf {\bibinfo {volume} {114}},\
  \bibinfo {pages} {1219} (\bibinfo {year} {1959})}\BibitemShut {NoStop}%
\bibitem [{\citenamefont {Levitt}(2008)}]{Levitt2008_Book}%
  \BibitemOpen
  \bibfield  {author} {\bibinfo {author} {\bibfnamefont {M.~H.}\ \bibnamefont
  {Levitt}},\ }\href@noop {} {\emph {\bibinfo {title} {Spin dynamics: basics of
  nuclear magnetic resonance}}}\ (\bibinfo  {publisher} {Wiley},\ \bibinfo
  {year} {2008})\BibitemShut {NoStop}%
\bibitem [{\citenamefont {Braginsky}\ and\ \citenamefont
  {Khalili}(1996)}]{Braginsky1996_RMP}%
  \BibitemOpen
  \bibfield  {author} {\bibinfo {author} {\bibfnamefont {V.~B.}\ \bibnamefont
  {Braginsky}}\ and\ \bibinfo {author} {\bibfnamefont {F.~Y.}\ \bibnamefont
  {Khalili}},\ }\href {\doibase 10.1103/RevModPhys.68.1} {\bibfield  {journal}
  {\bibinfo  {journal} {Rev. Mod. Phys.}\ }\textbf {\bibinfo {volume} {68}},\
  \bibinfo {pages} {1} (\bibinfo {year} {1996})}\BibitemShut {NoStop}%
\bibitem [{\citenamefont {Dementyev}\ \emph {et~al.}(2003)\citenamefont
  {Dementyev}, \citenamefont {Li}, \citenamefont {MacLean},\ and\ \citenamefont
  {Barrett}}]{Dementyev2003_PRB}%
  \BibitemOpen
  \bibfield  {author} {\bibinfo {author} {\bibfnamefont {A.~E.}\ \bibnamefont
  {Dementyev}}, \bibinfo {author} {\bibfnamefont {D.}~\bibnamefont {Li}},
  \bibinfo {author} {\bibfnamefont {K.}~\bibnamefont {MacLean}}, \ and\
  \bibinfo {author} {\bibfnamefont {S.~E.}\ \bibnamefont {Barrett}},\ }\href
  {\doibase 10.1103/PhysRevB.68.153302} {\bibfield  {journal} {\bibinfo
  {journal} {Phys. Rev. B}\ }\textbf {\bibinfo {volume} {68}},\ \bibinfo
  {pages} {153302} (\bibinfo {year} {2003})}\BibitemShut {NoStop}%
\bibitem [{\citenamefont {Hale}\ and\ \citenamefont
  {Mieher}(1969{\natexlab{b}})}]{Hale1969II_PR}%
  \BibitemOpen
  \bibfield  {author} {\bibinfo {author} {\bibfnamefont {E.~B.}\ \bibnamefont
  {Hale}}\ and\ \bibinfo {author} {\bibfnamefont {R.~L.}\ \bibnamefont
  {Mieher}},\ }\href {\doibase 10.1103/PhysRev.184.751} {\bibfield  {journal}
  {\bibinfo  {journal} {Phys. Rev.}\ }\textbf {\bibinfo {volume} {184}},\
  \bibinfo {pages} {751} (\bibinfo {year} {1969}{\natexlab{b}})}\BibitemShut
  {NoStop}%
\bibitem [{\citenamefont {Hale}\ and\ \citenamefont
  {Mieher}(1971)}]{Hale1971_PRB}%
  \BibitemOpen
  \bibfield  {author} {\bibinfo {author} {\bibfnamefont {E.~B.}\ \bibnamefont
  {Hale}}\ and\ \bibinfo {author} {\bibfnamefont {R.~L.}\ \bibnamefont
  {Mieher}},\ }\href {\doibase 10.1103/PhysRevB.3.1955} {\bibfield  {journal}
  {\bibinfo  {journal} {Phys. Rev. B}\ }\textbf {\bibinfo {volume} {3}},\
  \bibinfo {pages} {1955} (\bibinfo {year} {1971})}\BibitemShut {NoStop}%
\bibitem [{\citenamefont {Ivey}\ and\ \citenamefont
  {Mieher}(1972)}]{Ivey1972_PRL}%
  \BibitemOpen
  \bibfield  {author} {\bibinfo {author} {\bibfnamefont {J.~L.}\ \bibnamefont
  {Ivey}}\ and\ \bibinfo {author} {\bibfnamefont {R.~L.}\ \bibnamefont
  {Mieher}},\ }\href {\doibase 10.1103/PhysRevLett.29.176} {\bibfield
  {journal} {\bibinfo  {journal} {Phys. Rev. Lett.}\ }\textbf {\bibinfo
  {volume} {29}},\ \bibinfo {pages} {176} (\bibinfo {year} {1972})}\BibitemShut
  {NoStop}%
\bibitem [{\citenamefont {Ivey}\ and\ \citenamefont
  {Mieher}(1975{\natexlab{a}})}]{Ivey1975I_PRB}%
  \BibitemOpen
  \bibfield  {author} {\bibinfo {author} {\bibfnamefont {J.~L.}\ \bibnamefont
  {Ivey}}\ and\ \bibinfo {author} {\bibfnamefont {R.~L.}\ \bibnamefont
  {Mieher}},\ }\href {\doibase 10.1103/PhysRevB.11.822} {\bibfield  {journal}
  {\bibinfo  {journal} {Phys. Rev. B}\ }\textbf {\bibinfo {volume} {11}},\
  \bibinfo {pages} {822} (\bibinfo {year} {1975}{\natexlab{a}})}\BibitemShut
  {NoStop}%
\bibitem [{\citenamefont {Ivey}\ and\ \citenamefont
  {Mieher}(1975{\natexlab{b}})}]{Ivey1975II_PRB}%
  \BibitemOpen
  \bibfield  {author} {\bibinfo {author} {\bibfnamefont {J.~L.}\ \bibnamefont
  {Ivey}}\ and\ \bibinfo {author} {\bibfnamefont {R.~L.}\ \bibnamefont
  {Mieher}},\ }\href {\doibase 10.1103/PhysRevB.11.849} {\bibfield  {journal}
  {\bibinfo  {journal} {Phys. Rev. B}\ }\textbf {\bibinfo {volume} {11}},\
  \bibinfo {pages} {849} (\bibinfo {year} {1975}{\natexlab{b}})}\BibitemShut
  {NoStop}%
\bibitem [{\citenamefont {Mohiyaddin}\ \emph {et~al.}(2013)\citenamefont
  {Mohiyaddin}, \citenamefont {Rahman}, \citenamefont {Kalra}, \citenamefont
  {Klimeck}, \citenamefont {Hollenberg}, \citenamefont {Pla}, \citenamefont
  {Dzurak},\ and\ \citenamefont {Morello}}]{Mohiyaddin2013_NL}%
  \BibitemOpen
  \bibfield  {author} {\bibinfo {author} {\bibfnamefont {F.~A.}\ \bibnamefont
  {Mohiyaddin}}, \bibinfo {author} {\bibfnamefont {R.}~\bibnamefont {Rahman}},
  \bibinfo {author} {\bibfnamefont {R.}~\bibnamefont {Kalra}}, \bibinfo
  {author} {\bibfnamefont {G.}~\bibnamefont {Klimeck}}, \bibinfo {author}
  {\bibfnamefont {L.~C.~L.}\ \bibnamefont {Hollenberg}}, \bibinfo {author}
  {\bibfnamefont {J.~J.}\ \bibnamefont {Pla}}, \bibinfo {author} {\bibfnamefont
  {A.~S.}\ \bibnamefont {Dzurak}}, \ and\ \bibinfo {author} {\bibfnamefont
  {A.}~\bibnamefont {Morello}},\ }\href {\doibase 10.1021/nl303863s} {\bibfield
   {journal} {\bibinfo  {journal} {Nano Letters}\ }\textbf {\bibinfo {volume}
  {13}},\ \bibinfo {pages} {1903} (\bibinfo {year} {2013})}\BibitemShut
  {NoStop}%
\bibitem [{\citenamefont {Klimeck}\ \emph {et~al.}(2007)\citenamefont
  {Klimeck}, \citenamefont {Ahmed}, \citenamefont {Bae}, \citenamefont
  {Kharche}, \citenamefont {Clark}, \citenamefont {Haley}, \citenamefont {Lee},
  \citenamefont {Naumov}, \citenamefont {Ryu}, \citenamefont {Saied},
  \citenamefont {Prada}, \citenamefont {Korkusinski}, \citenamefont {Boykin},\
  and\ \citenamefont {Rahman}}]{Klimeck2007_IEEE}%
  \BibitemOpen
  \bibfield  {author} {\bibinfo {author} {\bibfnamefont {G.}~\bibnamefont
  {Klimeck}}, \bibinfo {author} {\bibfnamefont {S.}~\bibnamefont {Ahmed}},
  \bibinfo {author} {\bibfnamefont {H.}~\bibnamefont {Bae}}, \bibinfo {author}
  {\bibfnamefont {N.}~\bibnamefont {Kharche}}, \bibinfo {author} {\bibfnamefont
  {S.}~\bibnamefont {Clark}}, \bibinfo {author} {\bibfnamefont
  {B.}~\bibnamefont {Haley}}, \bibinfo {author} {\bibfnamefont
  {S.}~\bibnamefont {Lee}}, \bibinfo {author} {\bibfnamefont {M.}~\bibnamefont
  {Naumov}}, \bibinfo {author} {\bibfnamefont {H.}~\bibnamefont {Ryu}},
  \bibinfo {author} {\bibfnamefont {F.}~\bibnamefont {Saied}}, \bibinfo
  {author} {\bibfnamefont {M.}~\bibnamefont {Prada}}, \bibinfo {author}
  {\bibfnamefont {M.}~\bibnamefont {Korkusinski}}, \bibinfo {author}
  {\bibfnamefont {T.}~\bibnamefont {Boykin}}, \ and\ \bibinfo {author}
  {\bibfnamefont {R.}~\bibnamefont {Rahman}},\ }\href {\doibase
  10.1109/TED.2007.902879} {\bibfield  {journal} {\bibinfo  {journal} {Electron
  Devices, IEEE Transactions on}\ }\textbf {\bibinfo {volume} {54}},\ \bibinfo
  {pages} {2079} (\bibinfo {year} {2007})}\BibitemShut {NoStop}%
\bibitem [{\citenamefont {Muhonen}\ \emph {et~al.}(2014)\citenamefont
  {Muhonen}, \citenamefont {Dehollain}, \citenamefont {Laucht}, \citenamefont
  {Hudson}, \citenamefont {Sekiguchi}, \citenamefont {Itoh}, \citenamefont
  {Jamieson}, \citenamefont {McCallum}, \citenamefont {Dzurak},\ and\
  \citenamefont {Morello}}]{Muhonen2014}%
  \BibitemOpen
  \bibfield  {author} {\bibinfo {author} {\bibfnamefont {J.~T.}\ \bibnamefont
  {Muhonen}}, \bibinfo {author} {\bibfnamefont {J.~P.}\ \bibnamefont
  {Dehollain}}, \bibinfo {author} {\bibfnamefont {A.}~\bibnamefont {Laucht}},
  \bibinfo {author} {\bibfnamefont {F.~E.}\ \bibnamefont {Hudson}}, \bibinfo
  {author} {\bibfnamefont {T.}~\bibnamefont {Sekiguchi}}, \bibinfo {author}
  {\bibfnamefont {K.~M.}\ \bibnamefont {Itoh}}, \bibinfo {author}
  {\bibfnamefont {D.~N.}\ \bibnamefont {Jamieson}}, \bibinfo {author}
  {\bibfnamefont {J.~C.}\ \bibnamefont {McCallum}}, \bibinfo {author}
  {\bibfnamefont {A.~S.}\ \bibnamefont {Dzurak}}, \ and\ \bibinfo {author}
  {\bibfnamefont {A.}~\bibnamefont {Morello}},\ }\href@noop {} {\bibfield
  {journal} {\bibinfo  {journal} {arXiv:1402.7140}\ } (\bibinfo {year}
  {2014})}\BibitemShut {NoStop}%
\end{thebibliography}%

\end{document}